\documentstyle[epsfig,twocolumn,subfigure]{mn}

\newcommand{\goodgap}{\hspace{\subfigtopskip} \hspace{\subfigbottomskip}}

\title[GRBs afterglow and cosmological parameters]{Constraining cosmological parameters by Gamma Ray Burst X\,-\,ray afterglow lightcurves}
\author[V.F. Cardone et al.]{V.F. Cardone$^{1,2}$, M.G. Dainotti$^{3}$, S. Capozziello$^{2,4}$, R. Willingale$^{5}$ \\
$^1$ Dipartimento di Scienze e Tecnologie dell' Ambiente e del Territorio, Universit\`{a} degli Studi del Molise, \\
Contrada Fonte Lappone, 86090\,-\,Pesche (IS), Italy \\
$^2$Dipartimento di Scienze Fisiche, Universit\`{a} degli Studi di Napoli "Federico II", Complesso Universitario \\
di Monte Sant' Angelo, Edificio N, via Cinthia, 80126 - Napoli, Italy \\
$^3$Obserwatorium Astronomiczne, Uniwersytet Jagiello\'nski, ul. Orla 171, 31-501 Krak{\'o}w, Poland \\
$^4$I.N.F.N. - Sezione di Napoli, Complesso Universitario di Monte Sant'Angelo, Edificio 5, via Cinthia, 80126\,-\,Napoli, Italy \\
$^5$Department of Physics \& Astronomy, University of Leicester, Road Leicester LE1 7RH, United Kingdom \\}

\date{Accepted xxx, Received yyy, in original form zzz}

\begin{document}

\maketitle

\begin{abstract}

We present the Hubble diagram (HD) of 66 Gamma Ray Bursts (GRBs) derived using only data from their X\,-\,ray afterglow lightcurve. To this end, we use the recently updated $L_X$\,-\,$T_a$ correlation between the break time $T_a$ and the X\,-\,ray luminosity $L_X$ measured at $T_a$ calibrated from a sample of {\it Swift} GRBs with lightcurves well fitted by the Willingale et al. (2007) model. We then investigate the use of this HD to constrain cosmological parameters when used alone or in combination with other data showing that the use of GRBs leads to constraints in agreement with previous results in literature. We finally argue that a larger sample of high luminosity GRBs can provide a valuable information in the search for the correct cosmological model.

\end{abstract}

\maketitle

\section{Introduction}

That the universe is spatially flat, has a subcritical matter content and is undergoing a phase of accelerated expansion are nowadays globally accepted ideas. Strong evidence for this scenario come from the anisotropy and polarization spectra of the Cosmic Microwave Background Radiation (CMBR) \cite{Boom,QUAD,WMAP7}, the galaxy power spectrum \cite{Teg06,P07} with its Baryonic Acoustic Oscillations (BAO) \cite{Eis05,P10} and the Hubble diagram (HD) of Type Ia Supernovae (SNeIa) \cite{Union,H09,SNeIaSDSS}. What is driving this cosmic speed up and dominating the energy budget is, on the contrary, a still hotly debated question with answers running from the classical cosmological constant \cite{CPT92,SS00}, scalar fields \cite{PR03,Copeland} and higher order gravity theories \cite{CF08,FS08,ND08,dFT10}. In order to break the degeneracies among model parameters and go to the next level of model selection, it is mandatory to probe the background evolution up to very high redshift so that the onset of the acceleration epoch and the transition to the matter dominated regimes may be directly investigated.

\begin{figure*}
\centering
\subfigure{\includegraphics[width=7cm]{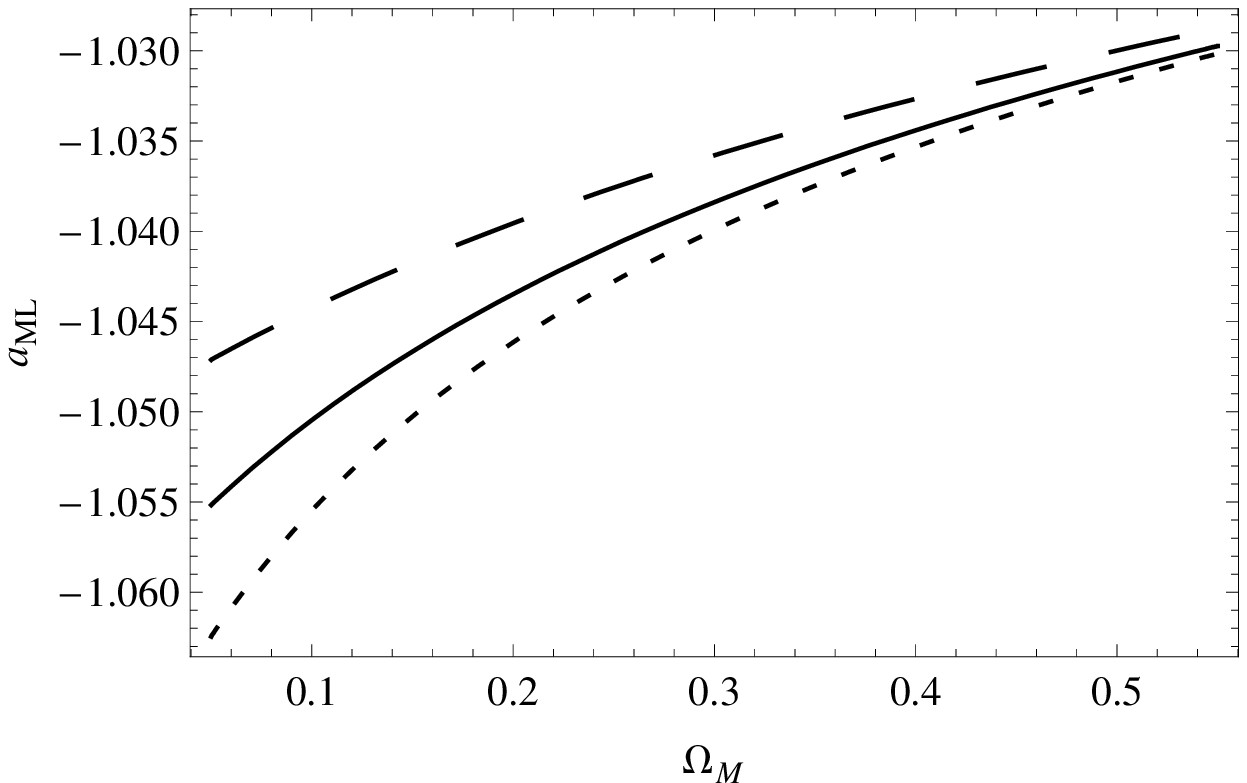}} \goodgap
\subfigure{\includegraphics[width=7cm]{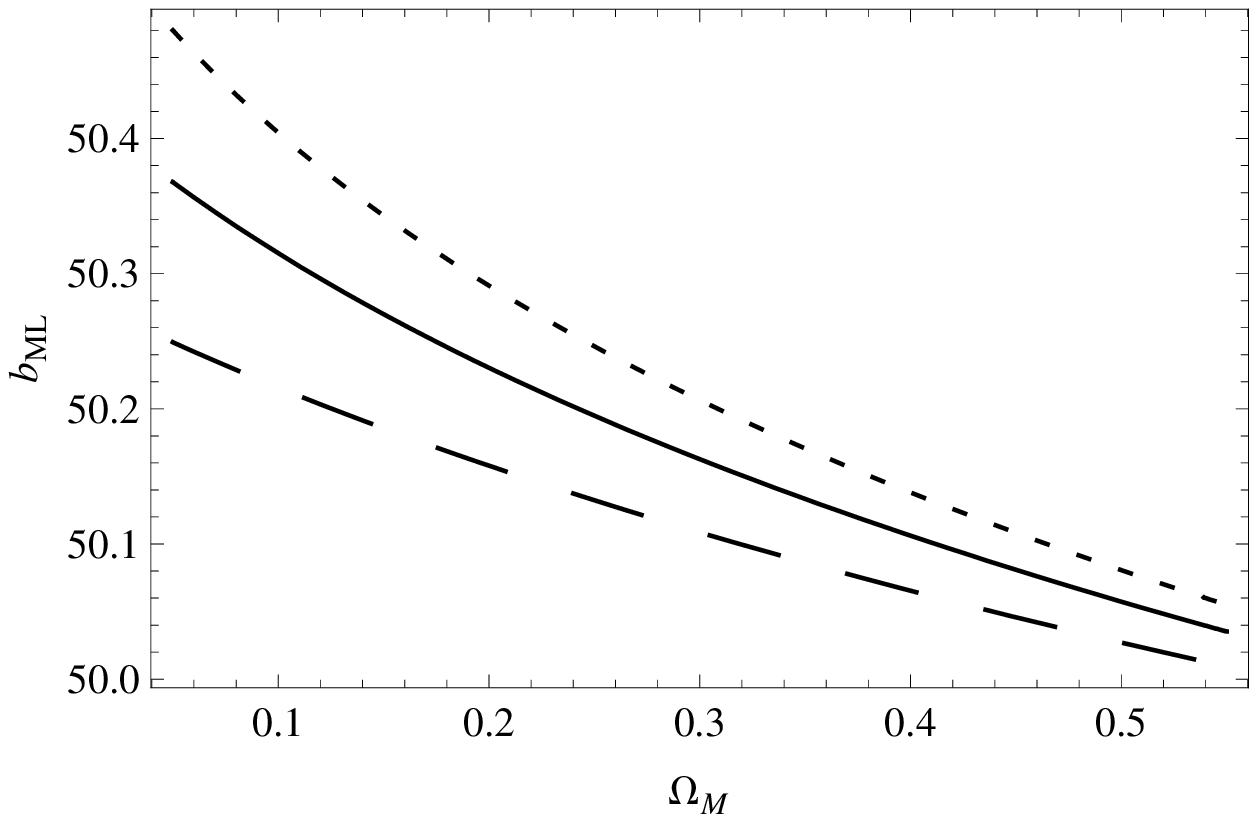}} \goodgap \\
\subfigure{\includegraphics[width=7cm]{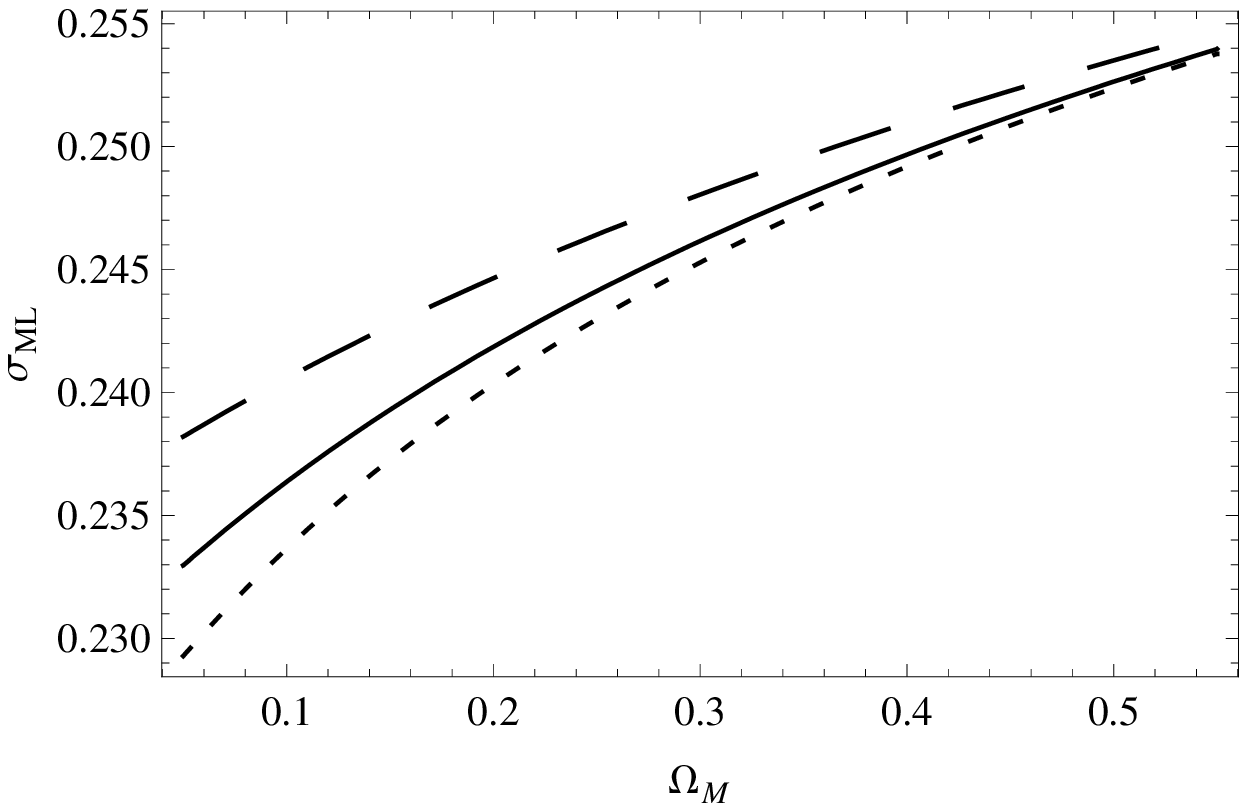}} \goodgap
\subfigure{\includegraphics[width=7cm]{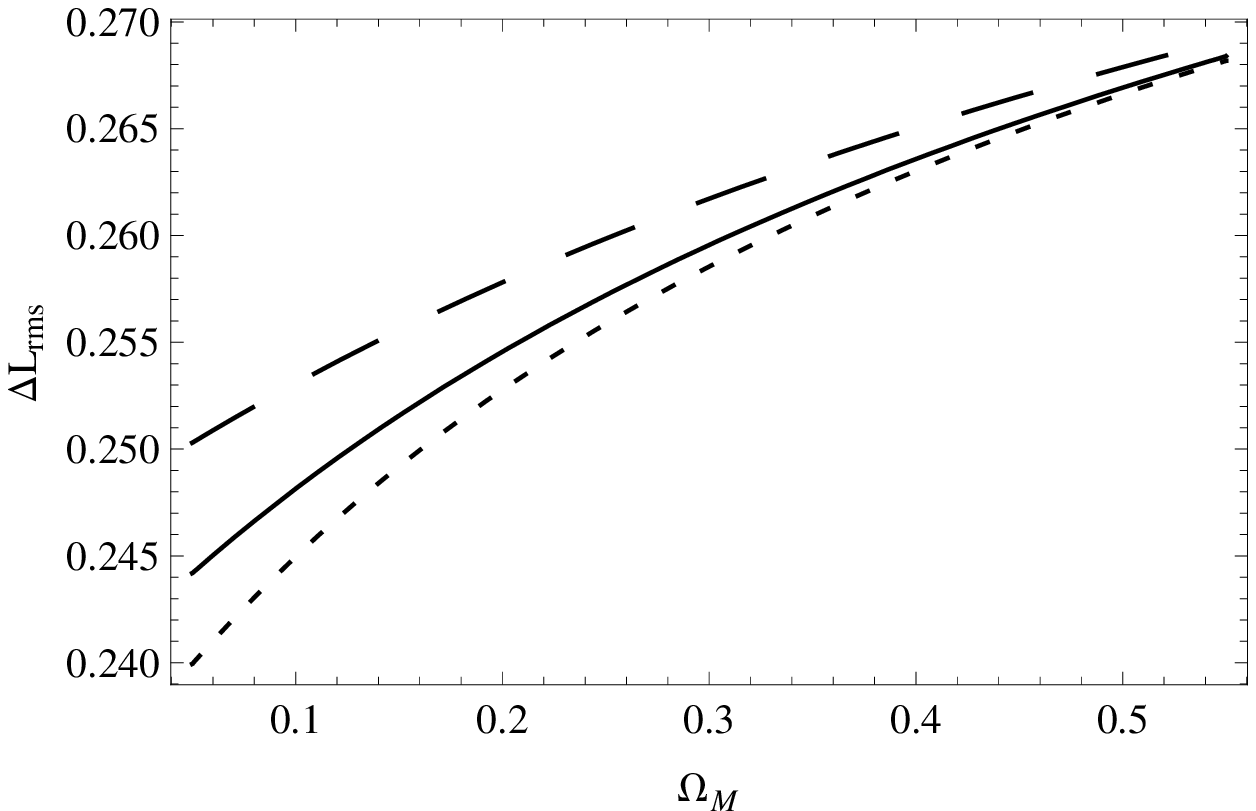}} \goodgap \\
\caption{Best fit calibration parameters $(a, b, \sigma_{int})$ and rms of the residuals for the $L_X$\,-\,$T_a$ correlation fitted to the 8 canonical GRBs as a function of the matter density parameter $\Omega_M$ for three different values of $w_0$, namely $w_0 = -1.25$ (short dashed), $w_0 = -1.0$ (solid), $w_0 = -0.75$ (long dashed). We use a CPL model with $h = 0.70$ and set $w_a = -w_0$ to have a matter dominated era at high redshift.}
\label{fig: cosmopar}
\end{figure*}

Thanks to their enormous energy release, GRBs are visible up to very high $z$, the largest one being at $z = 8.2$ \cite{Salvaterra2009}, hence appearing as ideal candidates for this task. Unfortunately, GRBs are everything but standard candles because their peak luminosity spans a wide range. There have nevertheless been many attempts to make them standardizeable candles resorting to the use of empirical correlations among distance dependent quantities and rest frame observables \cite{Amati08,FRR00,N000,G04,liza05}. Such empirical relations allow one to infer the GRB rest frame luminosity or energy from an observer frame measured quantity so that the distance modulus can be estimated with an error mainly depending on the intrinsic scatter of the adopted correlation. Combining the estimates from different correlations, Schaefer (2007) first derived the GRBs HD for 69 objects, while Cardone et al. (2009) used an enlarged sample and a different calibration method to update the GRBs HD. Many attempts on using GRBs as cosmological tools have since then been performed (see, e.g., Firmani et al. 2006, Liang et al. 2009, Qi \& Lu 2009, Izzo et al. 2009 and refs. therein) showing the interest in this application of GRBs.

In this {\it Letter}, we rely on the $L_X$\,-\,$T_a$ correlation \cite{D08} to build the HD of 66 GRBs from their X\,-\,ray afterglow lightcurves observed with the {\it Swift} satellite \cite{Evans2009}. We then present a preliminary application of the derived GRBs HD showing how this dataset can help constrain the parameters of some simple dark energy models.

\section{The $L_X$\,-\,$T_a$ correlation}

We use the updated version of the $L_X$\,-\,$T_a$ correlation between the luminosity $L_X$ at the break time $T_a$ and $T_a$ itself. We first remind the reader that $T_a$ is defined as the time marking the passage from the plateau phase to the power\,-\,law decay in the GRBs X\,-\,ray afterglow lightcurve as described by the universal fitting function proposed by Willingale et al. (2007, hereafter W07). Using a sample of 34 GRBs with lightcurve measured by the Swift satellite, Dainotti et al. (2008) first discovered that $L_X$ and $T_a$ are anticorrelated, as later confirmed by the semiempirical models of Ghisellini et al. (2009) and Yamazaki (2009). Recently, we have increased the GRBs sample and rederived the $L_X$\,-\,$T_a$ correlation (Dainotti et al. 2010, herefater D10). Introducing the error parameter $u = \sqrt{\sigma_{L_X}^2 + \sigma_{T_a}^2}$, D10 have also selected a class of high luminosity long GRBs with very well measured $(L_X, T_a)$ parameters ($u < 0.095$) and lightcurve closely matching the W07 model. Referring to this class of objects as {\it canonical} GRBs, D10 have demonstrated that they define an upper envelope for the $L_X$\,-\,$T_a$ correlation with the same slope, but a higher intercept than the one for the full sample. In order to avoid mixing objects possibly belonging to two different classes, we divide the 66 GRBs in two subsamples according to the value of $u$ being smaller or larger than 0.095. In the first case, we select 8 GRBs which we will refer to as the {\it canonical} (C) sample, while the remaining 58 will form the {\it non canonical} (NC) one. Adopting a flat $\Lambda$CDM model with $(\Omega_M, h) = (0.278, 0.699)$, we use the Bayesian method\footnote{The D' Agostini method provides a well motivated approach to deal with the problem of fitting a linear relation when the uncertainties on both the $(x, y)$ variable are comparable. Such a linear relation can be the outcome of a theoretical model so that one can expect that deviations from the underlying assumptions lead to the point scattering around the best fit line. This is what we refer to as {\it intrinsic scatter}. The D' Agostini method allows to take care of this term and estimate it in an unbiased way.} of D' Agostini (2005) to fit a linear relation, $\log{L_X} = a \log{[T_a/(1+z)]} + b$. We thus get\,:

\begin{equation}
\log{L_X} = -1.00 \log{\left ( \frac{T_a}{1 + z} \right )} + 49.26
\label{eq: lxtabfnc}
\end{equation}
with intrinsic scatter $\sigma_{int} = 0.66$ for the NC sample and

\begin{equation}
\log{L_X} = -1.04 \log{\left ( \frac{T_a}{1 + z} \right )} + 50.22
\label{eq: lxtabfc}
\end{equation}
with $\sigma_{int} = 0.23$ for the C sample. Note that these are the best fit values which have to be used in the later determination of the GRB distance modulus. However, the Bayesian approach allows to get the constraints on each one of the single $(a, b, \sigma_{int})$ by marginalizing over the other two. We thus find\,:

\begin{displaymath}
a = -1.00_{-0.21 \ -0.42}^{+0.21 \ +0.42} \ , \
b = 49.26_{-0.70 \ -1.37}^{+0.70 \ +1.41} \ , \
\end{displaymath}
\begin{displaymath}
\sigma_{int} = 0.70_{-0.11 \ -0.20}^{+0.13 \ +0.30} \ ,
\end{displaymath}
for the NC sample and

\begin{displaymath}
a = -1.04_{-0.22 \ -0.65}^{+0.23 \ +0.66} \ , \
b = 50.19_{-0.76 \ -1.50}^{+0.77 \ +1.55} \ , \
\end{displaymath}
\begin{displaymath}
\sigma_{int} = 0.41_{-0.16 \ -0.25}^{+0.38 \ +1.18}
\end{displaymath}
for the C sample, where we have reported the median value\footnote{Note that, because of parameters degeneracies, the marginalized likelihoods are not symmetric functions so that the median value may differ from the maximum likelihood one and the confidence ranges be asymmetric. See Dainotti et al. (2008) and refs. therein for a discussion of this issue.} and the $68$ and $95\%$ confidence ranges. In agreement with D09, we find that the slope is the same for the two samples, but the canonical GRBs are shifted to higher luminosities thus giving a larger zeropoint and defining an upper envelope for the $L_X$\,-\,$T_a$ correlation. Note that the intrinsic scatter for the canonical GRBs is much smaller thus leading to lower uncertainties on the estimated distance modulus. It is worth stressing that, although derived assuming a fiducial $\Lambda$CDM model, this has a negligible impact on the determination of the distance modulus \cite{CCD09}. To strengthen this result, we have determined the best fit parameters $(a, b, \sigma_{int})$ and the rms of the residuals for the C sample assuming a CPL \cite{CP01,L03} model and varying the parameters $(\Omega_M, w_0, h)$ while $w_a = -w_0$ (see later for the definition of these quantities). Fig.\,\ref{fig: cosmopar} shows that the best fit calibration parameters have a clear trend with $(\Omega_M, w_0)$, but the end to end variation of $(a, b)$, which enter the distance modulus estimate, is less than $3\%$. This is well within the uncertainties on the $(a, b)$ coefficients so that we are confident that the choice of the fiducial cosmological model used in the calibration procedure has a negligible impact on the derivation of the Hubble diagram. However, should future data allow us to increase the precision on $(a, b)$, one should likely readdress this problem looking for a model independent calibration.

\begin{figure}
\centering
\includegraphics[width=8cm]{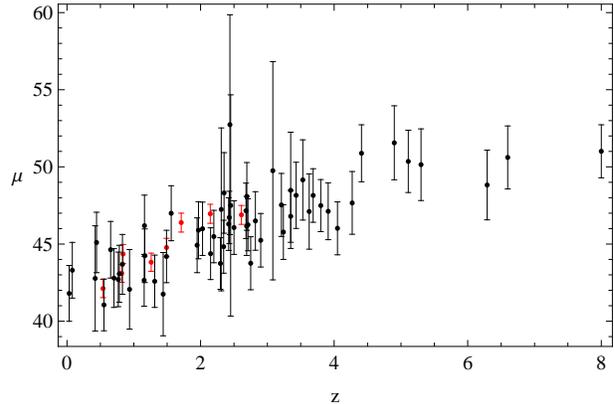}
\caption{GRBs Hubble diagram for the combined NC\,+\,C (black points) and C (red points) samples.}
\label{fig: grbhdplot}
\end{figure}

In order to infer the distance modulus of each GRB, we then simply note that $L_X$ is related to the luminosity distance $d_L(z)$ as \cite{D08,CCD09}\,:

\begin{equation}
L_X = 4 \pi d_L^2(z) (1 + z)^{-(2 + \beta)} F_X
\label{eq: lxdl}
\end{equation}
with $\beta$ the slope of the energy spectrum (modelled as a simple power\,-\,law) and $F_X$ the observed flux both measured at the break time $T_a$. Having measured $(T_a, \beta, F_X)$ and inferred $L_X$ using Eq.(\ref{eq: lxtabfnc}) or (\ref{eq: lxtabfc}), we can then estimate the GRB distance modulus as\,:

\begin{equation}
\mu(z) = 25 + 5 \log{d_L(z)} = 25 + \frac{5}{2} \log{\left [ \frac{L_X}{4 \pi (1 + z)^{-(2 + \beta)} F_X} \right ]}
\label{eq:_estmu}
\end{equation}
where $d_L(z)$ is in Mpc. The uncertainty is estimated by propagating the errors on $(\beta, F_X, L_X)$. Note that, when computing the error on $L_X$, we add in quadrature the uncertainty coming from $T_a$ and $\sigma_{int}$ to take care of the intrinsic scatter. When resorting to Eq.(4), we take care of the different calibration parameters for the two subsamples thus using Eq.(\ref{eq: lxtabfnc}) for the objects in the NC sample and Eq.(\ref{eq: lxtabfc}) for the C sample ones. The combined HD, shown in Fig.\,\ref{fig: grbhdplot}, covers the wide redshift range $(0.033, 8.2)$ thus showing that, although the error bars at the moment are still quite large, the $L_X$\,-\,$T_a$ correlation could allow us to probe of both the dark energy epoch and the matter dominated era with a single tracer.

\begin{table*}
\caption{Constraints on the model parameters using GRBs and priors on $(h, \omega_M)$. Columns are as follow\,: 1. id of the model, 2. maximum likelihood parameters, 3.,4.,5.,6. median and 68 and 95$\%$ CL after marginalization. Upper (lower) half of the table refers to the results using the NC + C (C only) sample. A sign "-" means that the parameter is fixed to its theoretical value.}
\begin{center}
\begin{tabular}{|c|c|c|c|c|c|}
\hline
Id & ${\bf p}_{ML}$ & $\Omega_M$ & $w_0$ & $w_a$ & $h$ \\
\hline \hline

~ & ~ & ~ & ~ & ~ & ~ \\

$\Lambda$CDM & $(0.248, 0.739)$ & $0.246_{-0.024 \ -0.039}^{+0.022 \ +0.050}$ & --- & --- &
$0.742_{-0.032 \ -0.066}^{+0.040 \ +0.068}$ \\

~ & ~ & ~ & ~ & ~ & ~ \\

QCDM & $(0.248, -0.37, 0.741)$ & $0.248_{-0.024 \ -0.042}^{+0.026 \ +0.052}$ & $-0.52_{-0.38 \ -0.82}^{+0.14 \ +0.19}$ & --- & $0.739_{-0.034 \ -0.067}^{+0.035 \ +0.067}$ \\

~ & ~ & ~ & ~ & ~ & ~ \\

CPL & $(0.242, -0.34, 0.02, 0.751)$ & $0.247_{-0.023 \ -0.043}^{+0.029 \ +0.066}$ & $-0.48_{-0.38 \ -0.96}^{+0.11 \ +0.14}$ & $-0.61_{-1.39 \ -2.21}^{+1.20 \ +2.43}$ & $0.740_{-0.038 \ -0.080}^{+0.037 \ +0.070}$ \\

~ & ~ & ~ & ~ & ~ & ~ \\
\hline
~ & ~ & ~ & ~ & ~ & ~ \\

$\Lambda$CDM & $(0.246, 0.743)$ & $0.246_{-0.020 \ -0.037}^{+0.027 \ +0.057}$ & --- & --- &
$0.743_{-0.039 \ -0.067}^{+0.031 \ +0.061}$ \\

~ & ~ & ~ & ~ & ~ & ~ \\

QCDM & $(0.244, -1.26, 0.745)$ & $0.246_{-0.024 \ -0.041}^{+0.026 \ +0.055}$ & $-0.73_{-0.72 \ -0.91}^{+0.35 \ +0.40}$ & --- & $0.743_{-0.036 \ -0.069}^{+0.036 \ +0.074}$ \\

~ & ~ & ~ & ~ & ~ & ~ \\
CPL & $(0.245, -0.36, -2.89, 0.744)$ & $0.242_{-0.022 \ -0.046}^{+0.028 \ +0.064}$ & $-0.73_{-0.43 \ -0.85}^{+0.27 \ +0.38}$ & $0.01_{-1.90 \ -2.82}^{+1.44 \ +2.55}$ & $0.748_{-0.036 \ -0.080}^{+0.034 \ +0.077}$ \\

~ & ~ & ~ & ~ & ~ & ~ \\

\hline
\end{tabular}
\end{center}
\end{table*}

\section{GRBs HD as a cosmological tool}

Whatever is the tracer used, the HD is a primary tool to investigate the viability of a cosmological model. Indeed, the luminosity distance $d_L(z)$ reads\,:

\begin{equation}
d_L(z) = \frac{c}{H_0} (1 + z) \int_{0}^{z}{\frac{dz'}{E(z')}}
\label{eq: defdl}
\end{equation}
with $E(z) = H(z)/H_0$ the dimensionless Hubble parameter. For a spatially flat universe made out of dust matter and dark energy with the CPL equation of state (eos) $w(z) = w_0 + w_a z/(1 + z)$, it is\,:

\begin{equation}
E^2 = \Omega_M (1 + z)^3 + (1 - \Omega_M) (1 + z)^{3(1 + w_0 + w_a)} \exp{\left (
- \frac{3 w_a z}{1 + z} \right )}
\label{eq: ezcpl}
\end{equation}
where $\Omega_M$ is the present day matter density parameter and one recovers the $\Lambda$CDM model for $(w_0, w_a) = (-1, 0)$. In order to constrain the model parameters, we use a Markov Chain Monte Carlo (MCMC) algorithm to maximize the likelihood function ${\cal{L}}({\bf p}) \propto \exp{[-\chi^2({\bf p})/2]}$ where ${\bf p}$ is the set of model parameters and the expression for $\chi^2({\bf p})$ depends on the dataset used. Since we are here interested in testing the usefulness of the $L_X$\,-\,$T_a$ correlation as a cosmological tool, we consider, as a first test, GRBs only thus setting\,:

\begin{eqnarray}
\chi^2({\bf p}) & = & \sum_{i = 1}^{{\cal{N}}_{GRB}}{\left [ \frac{\mu_{obs}(z_i) - \mu_{th}(z_i, {\bf p})}{\sigma_i} \right ]^2} \nonumber \\
& + &  \left ( \frac{h - 0.742}{0.036} \right )^2
+ \left ( \frac{\omega_M - 0.1356}{0.0034} \right )^2 \ .
\label{eq: defchigrbonly}
\end{eqnarray}
Here, $\mu_{obs}$ and $\mu_{th}$ are the observed and theoretically predicted values of the distance modulus, while the sum is over the ${\cal{N}}_{GRB}$ GRBs in the sample. The last two terms are Gaussian priors on $h$ and $\omega_M = \Omega_M h^2$ and are included in order to help break the degeneracies among model parameters. To this aim, we have resorted to the results of the SHOES collaboration \cite{shoes} and the WMAP7 constraints \cite{WMAP7}, respectively, to set the numbers used in Eq.(\ref{eq: defchigrbonly}). We consider three different cosmological models, namely the $\Lambda$CDM one, quiessence (QCDM) obtained by setting $w_a = 0$ but $w_0$ free to vary, and the full CPL model leaving both $(w_0, w_a)$ unspecified. The results of the MCMC analysis are  summarized in Table 1 for the three models considered and two GRBs samples, namely the full NC\,+\,C set or the C subsample only. Note that this test allows us to investigate whether a sample made of only high precision GRBs is better suited to constrain  cosmological parameters. As a general remark, we note that the constraints are in good agreement with previous ones in literature \cite{Union,H09,WMAP7,P10} thus showing that using GRBs does not introduce any bias in the estimate of cosmological parameters. While the constraints on $(\Omega_M, h)$ are quite narrow, those on the eos parameters are quite weak\footnote{Note that the $68\%$ and $95\%$ CL on $w_0$ in Table 1 are essentially driven by the assumption $w_0 \le -1/3$ we have imposed in order to have $\rho + 3 p \le 0$ for the dark energy fluid.} when deviations from the $\Lambda$CDM model are allowed. This is not surprising since the GRBs HDs mainly probe the matter dominated high\,-\,$z$ regime. As such, the details of the eos have only a minor impact on the background evolution and the HD. We nevertheless note that, although with a large error, the fit to the CPL model points towards $w_a = 0$, i.e. a constant eos. Considering that $w_0 = -1$ is well within the $68\%$ CL range, one can argue that the $\Lambda$CDM model is still preferred by a GRBs only fit. Such a conclusion is also strengthened noting that the reduced $\chi^2$ (i.e., $\chi^2/({\cal{N}}_{GRB} - n_p)$ with $n_p$ the number of parameters) for the maximum likelihood model is the same for the three cases considered so that increasing $n_p$ is not statistically supported.

It is worth comparing the results from the NC\,+\,C and C samples. The canonical GRBs cover a narrower redshift range ($0.125 \le z \le 2.612$ (with a median $z_{med} = 1.26$) than the NC sample (with $0.033 \le z \le 8.2$ and $z_{med} = 2.31$), but with significantly less uncertainties on the distance modulus estimates. In a sense, one can say that comparing the results from the two samples helps understanding which strategy (increasing statistics or ameliorating precision) should be chosen to improve the constraint on cosmological parameters by GRBs. Comparing the values in the upper and lower half of Table 1, one sees that the percentage error on each parameter is comparable using either the NC\,+\,C or the C sample, while the constraints on the parameters (in particular, $\Omega_M$ and $h$) are in remarkable good agreement. One can therefore argue that the reduction in the number of GRBs from the NC\,+\,C to the C sample has been compensated by the increased accuracy, but a more extensive analysis have to be carried out before a definitive answer could be drawn.

\begin{table*}
\caption{Same as Table 1 but using GRBs + SHOES + BAO + ${\cal{R}}$ + $H(z)$.}
\begin{center}
\begin{tabular}{|c|c|c|c|c|c|}
\hline
Id & ${\bf p}_{ML}$ & $\Omega_M$ & $w_0$ & $w_a$ & $h$ \\
\hline \hline

~ & ~ & ~ & ~ & ~ & ~ \\

$\Lambda$CDM & $(0.265, 0.709)$ & $0.263_{-0.021 \ -0.041}^{+0.026 \ +0.054}$ & --- & --- &
$0.709_{-0.014 \ -0.030}^{+0.015 \ +0.034}$ \\

~ & ~ & ~ & ~ & ~ & ~ \\

QCDM & $(0.271, -0.98, 0.707)$ & $0.267_{-0.032 \ -0.061}^{+0.035 \ +0.072}$ & $-1.00_{-0.15 \ -0.37}^{+0.14 \ +0.29}$ & --- & $0.710_{-0.025 \ -0.051}^{+0.025 \ +0.057}$ \\

~ & ~ & ~ & ~ & ~ & ~ \\

CPL & $(0.262, -1.32, 1.01, 0.726)$ & $0.273_{-0.036 \ -0.064}^{+0.036 \ +0.069}$ & $-1.01_{-0.33 \ -0.54}^{+0.47 \ +0.64}$ & $-0.02_{-1.45 \ -2.44}^{+0.88 \ +1.57}$ & $0.706_{-0.032 \ -0.053}^{+0.033 \ +0.062}$ \\

~ & ~ & ~ & ~ & ~ & ~ \\
\hline
~ & ~ & ~ & ~ & ~ & ~ \\

$\Lambda$CDM & $(0.262, 0.710)$ & $0.263_{-0.024 \ -0.048}^{+0.022 \ +0.049}$ & ---- & ---- &
$0.710_{-0.016 \ -0.031}^{+0.016 \ +0.028}$ \\

~ & ~ & ~ & ~ & ~ & ~ \\

QCDM & $(0.244, -1.26, 0.745)$ & $0.246_{-0.024 \ -0.041}^{+0.026 \ +0.055}$ & $-0.73_{-0.71 \ -0.91}^{+0.35 \ +0.40}$ & ---- & $0.743_{-0.035 \ -0.069}^{+0.037 \ +0.074}$ \\

~ & ~ & ~ & ~ & ~ & ~ \\

CPL & $(0.249, -1.50, 1.40, 0.742)$ & $0.256_{-0.032 \ -0.060}^{+0.037 \ +0.085}$ & $-1.06_{-0.38 \ -0.54}^{+0.43 \ +0.64}$ & $-0.02_{-1.35 \ -2.49}^{+1.06 \ +1.54}$ & $0.717_{-0.034 \ -0.064}^{+0.033 \ +0.059}$ \\

~ & ~ & ~ & ~ & ~ & ~ \\

\hline
\end{tabular}
\end{center}
\end{table*}

As a next step, we combine the GRBs HDs with other data redefining ${\cal{L}}({\bf p})$ as\,:

\begin{eqnarray}
{\cal{L}}({\bf p}) & \propto & \frac{\exp{(-\chi^2_{GRB}/2})}{(2 \pi)^{{\cal{N}}_{GRB}/2} |{\bf C}_{GRB}|^{1/2}} \nonumber \\
~ & \times  & \frac{1}{\sqrt{2 \pi \sigma_h^2}} \exp{\left [ - \frac{1}{2} \left ( \frac{h - h_{obs}}{\sigma_h} \right )^2
\right ]} \nonumber \\
~ & \times & \frac{\exp{(-\chi^2_{BAO}/2})}{(2 \pi)^{{\cal{N}}_{BAO}/2} |{\bf C}_{BAO}|^{1/2}} \nonumber \\
~ & \times & \frac{1}{\sqrt{2 \pi \sigma_{{\cal{R}}}^2}} \exp{\left [ - \frac{1}{2} \left ( \frac{{\cal{R}} - {\cal{R}}_{obs}}{\sigma_{{\cal{R}}}} \right )^2 \right ]} \nonumber \\
~ & \times & \frac{\exp{(-\chi^2_{H}/2})}{(2 \pi)^{{\cal{N}}_{H}/2} |{\bf C}_{H}|^{1/2}} \  .
\label{eq: defchiall}
\end{eqnarray}
Here, the first two terms are the same as in Eq.(\ref{eq: defchigrbonly}) with ${\bf C}_{GRB}$ the GRBs diagonal covariance matrix and $(h_{obs}, \sigma_h) = (0.742, 0.036)$. The third term takes into account the constraints on $d_z = r_s(z_d)/D_V(z)$ with $r_s(z_d)$ the comoving sound horizon at the drag redshift $z_d$ (which we fix to be $r_s(z_d) = 152.6 \ {\rm Mpc}$ from WMAP7) and the volume distance is defined as \cite{Eis05}\,:

\begin{equation}
D_V(z) = \left \{ \frac{c z}{H(z)} \left [ \frac{D_L(z)}{1 + z} \right ]^2 \right \}^{1/3} \ .
\label{eq: defdv}
\end{equation}
The values of $d_z$ at $z = 0.20$ and $z = 0.35$ have been estimated by Percival et al. (2010) using the SDSS DR7 galaxy sample so that we define $\chi^2_{BAO} = {\bf D}^T {\bf C}_{BAO}^{-1} {\bf C}$ with ${\bf D}^T = (d_{0.2}^{obs} - d_{0.2}^{th}, d_{0.35}^{obs} - d_{0.35}^{th})$ and ${\bf C}_{BAO}$ the BAO covariance matrix. The next term refers to the shift parameter \cite{B97,EB99}\,:

\begin{equation}
{\cal{R}} = \sqrt{\Omega_M} \int_{0}^{z_{\star}}{\frac{dz'}{E(z')}}
\label{eq: defshiftpar}
\end{equation}
with $z_\star = 1090.10$ the redshift of the last scattering surface. We follow again WMAP7 setting $({\cal{R}}_{obs}, \sigma_{{\cal{R}}}) = (1.725, 0.019)$. While all these quantities (except for the Gaussian prior on $h$) mainly involve the integrated $E(z)$, the last term refers to the actual measurements of $H(z)$ from the differential age of passively evolving elliptical galaxies. We then use the data collected by Stern et al. (2010) giving the values of the Hubble parameter for ${\cal{N}}_H = 11$ different points over the redshift range $0.10 \le z \le 1.75$ with a diagonal covariance matrix. The results obtained fitting this combined dataset are summarized in Table 2. Not surprisingly, we find that the $\Lambda$CDM model is still statistically favoured by the combined dataset. The constraints on $(\Omega_M, h)$ for this case are almost unchanged by the addition of the other data. This is expected since the data we have added help break the degeneracy between the matter content and the eos parameters. Since in the $\Lambda$CDM case the eos is set from the beginning, the priors on $(h, \omega_M)$ we have used to get Table 1 are essentially equivalent to the full dataset in Table 2. The situation is different for the QCDM and CPL models since the eos is allowed to vary. The combined dataset is now able to constrain $(w_0, w_a)$ thanks to the $H(z)$ and BAO data which probe the redshift range where dark energy drives the background evolution thus complementing GRBs. In order to better show the impact of GRBs, we have repeated the fit excluding their HD. Considering, as a test case, the CPL model, we find for the maximum likelihood parameters $(\Omega_M, w_0, w_a, h) = (0.258, -1.52, 1.51, 0.739)$ close to that obtained using the GRBs. The marginalized constraints now read\,:

\begin{displaymath}
\Omega_M = 0.263_{-0.034 \ -0.061}^{+0.033 \ +0.062} \ , \
w_0 = -1.00_{-0.42 \ -0.60}^{+0.35 \ +0.56} \ , \
\end{displaymath}
\begin{displaymath}
w_a = -0.22_{-1.15 \ -2.29}^{+1.30 \ +1.82} \ , \
h = 0.714_{-0.029 \ -0.053}^{+0.031 \ +0.056} \ .
\end{displaymath}
Comparing these values to those in Table 2 shows that adding GRBs does not significantly narrow the parameters confidence ranges. This is actually expected since the present GRBs datasets are affected by large errors (the NC sample) or few statistics (the C sample). However, it is worth noting that GRBs help push the constraints on $w_a$ towards 0 thus suggesting that the inclusion of a large sample of canonical ($u < 0.095$) GRBs may strengthen the case for a constant eos dark energy model.

Up to now, we have not used SNeIa since we have been mainly interested in investigating how GRBs (and the $L_X$\,-\,$T_a$ correlation) could be used as cosmological tools. Having demonstrated that the inclusion of GRBs does not bias the search for cosmological parameters, we finally add SNeIa to the above combined dataset using the Constitution sample \cite{H09} comprising 397 objects over the range $0.015 \le z \le 1.551$. For the CPL model and the NC\,+\,C GRBs, we get $(\Omega_M, w_0, w_a, h) = (0.279, -1.00, 0.16, 0.708)$ as maximum likelihood parameters, while the marginalized constraints on the single quantities are as follows\,:

\begin{displaymath}
\Omega_M = 0.278_{-0.022 \ -0.041}^{+0.021 \ +0.054} \ , \
w_0 = -1.00_{-0.17 \ -0.30}^{+0.15 \ +0.32} \ , \
\end{displaymath}
\begin{displaymath}
w_a = 0.09_{-0.62 \ -1.69}^{+0.68 \ +1.16} \ , \
h = 0.703_{-0.015 \ -0.032}^{+0.019 \ +0.035} \ .
\end{displaymath}
Not surprisingly, the maximum likelihood model is close to the $\Lambda$CDM $(w_0, w_a) = (-1, 0)$ case, while the constraints on the parameters are in full agreement with those reported in Tables 1 and 2 being narrower thanks to the SNeIa data. In particular, the eos parameters are now better constrained since the SNeIa HD mainly probes the low redshift dark energy dominated universe. We have not tried to repeat this analysis using the C sample since, due to their high number and good accuracy, SNeIa actually dominate the fit as can be checked by fitting SNeIa only without the inclusion of GRBs or the other data.

\section{Conclusions}

A {\it ground zero} test of every cosmological model is its ability to fit the observed Hubble diagram. Such a test tells us whether the model is able to give rise to the observed background evolution and select the region of the parameter space leading to the correct sequence of accelerating and decelerating expansion. Although still being the primary tracer of the Hubble diagram, SNeIa may trace cosmic evolution only up to $z \sim 2$ so that it is mandatory to look for a different class of astronomical objects to go beyond this limit and investigate the matter dominated era. GRBs are considered the ideal candidates for this role thus motivating the hunt for a calibration method to make them standardizeable candles. We have presented here the use of the $L_X$\,-\,$T_a$ correlation as a valid method to infer the GRBs HD up to $z = 8.2$ using data coming from the X\,-\,ray lightcurve. It is worth stressing that this is the only empirical law relating quantities measured from the afterglow lightcurve rather than being related to the prompt emission quantities. Moreover, differently from what has yet been done in the past \cite{S07,CCD09}, the HD for the NC\,+\,C sample is the only GRBs HD based on the use of a single correlation and containing a statistically meaningful number of objects. The use of the $L_X$\,-\,$T_a$ correlation then avoids the need of combining different correlations to increase the number of GRBs with a known distance modulus. Each correlation is affected by its own possible systematics and characterized by different intrinsic scatter so that combining all of them in a single HD can introduce unexpected features and hence bias the constraints on the cosmological parameters.

The intrinsic scatter of the $L_X$\,-\,$T_a$ correlation may be significantly reduced if one considers only its upper envelope defined by the canonical GRBs. On the other hand, using the full 66 GRBs, we have been able to get constraints on the matter content and the present day Hubble constant comparable to those yet available in literature. Such a preliminary investigation has convincingly shown that the $L_X$\,-\,$T_a$ correlation may be used to construct a GRBs HD which does not introduce any bias in the search for cosmological parameters. It is worth stressing that the same results have been obtained using the canonical GRBs despite the fact that they represent just $\sim 12\%$ of the full sample. We therefore argue that an observational effort dedicated to look for canonical GRBs may turn them from ideal candidates to actual tools for efficiently investigating the dark energy puzzle. \\

{\it Acknowledgements.} We warmly thank M. Ostrowski for his valuable comments. This work made use of data supplied by the UK Swift Science Data Centre at the University of Leicester. MGD is grateful for the support from Polish MNiSW through the grant N N203 380336. MGD is also grateful for the support from Angelo Della Riccia Foundation.

\end{document}